\documentclass[twoside,twocolumn,9pt]{article}

\usepackage[usetitle,usedoi,linkdoi]{rsc}

\usepackage{extsizes}
\usepackage[version=3]{mhchem}
\usepackage[left=1.5cm, right=1.5cm, top=1.785cm, bottom=2.0cm]{geometry}
\usepackage{balance}
\usepackage{times,mathptmx}
\usepackage{sectsty}
\usepackage{graphicx} 
\usepackage{lastpage}
\usepackage[format=plain,justification=justified,singlelinecheck=false,font={stretch=1.125,small,sf},labelfont=bf,labelsep=space]{caption}
\usepackage{float}
\usepackage{fancyhdr}
\usepackage{fnpos}
\usepackage[english]{babel}
\addto{\captionsenglish}{%
  
}
\usepackage{array}
\usepackage{droidsans}
\usepackage{charter}
\usepackage[T1]{fontenc}
\usepackage[usenames,dvipsnames]{xcolor}
\usepackage{setspace}
\usepackage[compact]{titlesec}
\usepackage{hyperref}
  \hypersetup{colorlinks = true, 	% colored links
  citecolor = blue,		% blue links for refs
  linkcolor = blue, 		% red links for figs
  breaklinks = true,
  backref = true			% enable going back from link target
  }

\usepackage{chemformula}
\usepackage[space]{grffile} %file names with spaces
\usepackage{amssymb}

\newcommand*{\f}{\bfseries}                      %bold
\newcommand*{\n}{\mdseries \upshape \rmfamily}    %back to standard
\newcommand*{\grad}{$^\circ$}
\newcommand*{\gradC}{\,$^\circ$C}
\newcommand*{\zro}{\ce{ZrO2}}
\newcommand*{\ptzr}{\ce{Pt3Zr}}
\newcommand*{\pox}{$p_\mathrm{O_2}$}
\newcommand*{\muox}{$\mu_\mathrm{\nicefrac{1}{2} O_2}$}
\newcommand*{\ox}{O$_2$}

\newcommand*{\p}[2]{$#1 \times 10^{-#2}$\,mbar}
\newcommand*{\EB}{$E_\mathrm{B}$}
\newcommand*{\onebyone}{($1 \times 1$)}

\newcommand*{\twobyone}{($2 \times 1$)}
\newcommand*{\sn}{($\sqrt{19} \times \sqrt{19}$)}

\newcommand*\latin[1]{\textit{#1}}

\definecolor{cream}{RGB}{222,217,201}

\begin{document}

%%%%%%%%%%%%%%%%%%%%%%%%%%%%%%%%%%%%%%%%%%%%%%%%%%%%%%%%%%%%%%%%%%%%%
%% Header
%%%%%%%%%%%%%%%%%%%%%%%%%%%%%%%%%%%%%%%%%%%%%%%%%%%%%%%%%%%%%%%%%%%%%

\pagestyle{fancy}
\thispagestyle{plain}
\fancypagestyle{plain}{

%%%HEADER%%%
%\fancyhead[C]{\includegraphics[width=18.5cm]{head_foot/header_bar}}
%\fancyhead[L]{\hspace{0cm}\vspace{1.5cm}\includegraphics[height=30pt]{head_foot/journal_name}}
%\fancyhead[R]{\hspace{0cm}\vspace{1.7cm}\includegraphics[height=55pt]{head_foot/RSC_LOGO_CMYK}}
\renewcommand{\headrulewidth}{0pt}
}
%%%END OF HEADER%%%

%%%PAGE SETUP - Please do not change any commands within this section%%%
\makeFNbottom
\makeatletter
\renewcommand\LARGE{\@setfontsize\LARGE{15pt}{17}}
\renewcommand\Large{\@setfontsize\Large{12pt}{14}}
\renewcommand\large{\@setfontsize\large{10pt}{12}}
\renewcommand\footnotesize{\@setfontsize\footnotesize{7pt}{10}}
\makeatother

\renewcommand{\thefootnote}{\fnsymbol{footnote}}
\renewcommand\footnoterule{\vspace*{1pt}% 
\color{cream}\hrule width 3.5in height 0.4pt \color{black}\vspace*{5pt}} 
\setcounter{secnumdepth}{5}

\makeatletter 
\renewcommand\@biblabel[1]{#1}            
\renewcommand\@makefntext[1]% 
{\noindent\makebox[0pt][r]{\@thefnmark\,}#1}
\makeatother 
\renewcommand{\figurename}{\small{Fig.}~}
\sectionfont{\sffamily\Large}
\subsectionfont{\normalsize}
\subsubsectionfont{\bf}
\setstretch{1.125} %In particular, please do not alter this line.
\setlength{\skip\footins}{0.8cm}
\setlength{\footnotesep}{0.25cm}
\setlength{\jot}{10pt}
\titlespacing*{\section}{0pt}{4pt}{4pt}
\titlespacing*{\subsection}{0pt}{15pt}{1pt}
%%%END OF PAGE SETUP%%%

%%%FOOTER%%%
\fancyfoot{}
%\fancyfoot[LO,RE]{\vspace{-7.1pt}\includegraphics[height=9pt]{head_foot/LF}}
%\fancyfoot[CO]{\vspace{-7.1pt}\hspace{13.2cm}\includegraphics{head_foot/RF}}
%\fancyfoot[CE]{\vspace{-7.2pt}\hspace{-14.2cm}\includegraphics{head_foot/RF}}
\fancyfoot[RO]{\footnotesize{\sffamily{1--\pageref{LastPage} ~\textbar  \hspace{2pt}\thepage}}}
\fancyfoot[LE]{\footnotesize{\sffamily{\thepage~\textbar\hspace{3.45cm} 1--\pageref{LastPage}}}}
\fancyhead{}
\renewcommand{\headrulewidth}{0pt} 
\renewcommand{\footrulewidth}{0pt}
\setlength{\arrayrulewidth}{1pt}
\setlength{\columnsep}{6.5mm}
\setlength\bibsep{1pt}
%%%END OF FOOTER%%%

%%%FIGURE SETUP - please do not change any commands within this section%%%
\makeatletter 
\newlength{\figrulesep} 
\setlength{\figrulesep}{0.5\textfloatsep} 

\newcommand{\topfigrule}{\vspace*{-1pt}% 
\noindent{\color{cream}\rule[-\figrulesep]{\columnwidth}{1.5pt}} }

\newcommand{\botfigrule}{\vspace*{-2pt}% 
\noindent{\color{cream}\rule[\figrulesep]{\columnwidth}{1.5pt}} }

\newcommand{\dblfigrule}{\vspace*{-1pt}% 
\noindent{\color{cream}\rule[-\figrulesep]{\textwidth}{1.5pt}} }

\makeatother
%%%END OF FIGURE SETUP%%%

%%%%%%%%%%%%%%%%%%%%%%%%%%%%%%%%%%%%%%%%%%%%%%%%%%%%%%%%%%%%%%%%%%%%%
%% Title, Author and Abstract
%%%%%%%%%%%%%%%%%%%%%%%%%%%%%%%%%%%%%%%%%%%%%%%%%%%%%%%%%%%%%%%%%%%%%

\twocolumn[
  \begin{@twocolumnfalse}
\vspace{3cm}
\sffamily
\begin{tabular}{m{4.5cm} p{13.5cm} }

{\footnotesize Submitted to J. Mater. Chem. A}
& \noindent\LARGE{\textbf{Substoichiometric ultrathin zirconia films cause strong metal-support interaction$^\dag$}} \\
%  Oxygen spillover on zirconia thin films observed via XPS energy levels
\vspace{0.3cm} & \vspace{0.3cm} \\

 & \noindent\large{
 Peter Lackner{$^{a}$},
 Joong-Il Jake Choi{$^{a,\ddag}$},
 Ulrike Diebold{$^{a}$},
 and Michael Schmid{$^{a,\ast}$}} \\

{\footnotesize 2019-Aug-02}
& \noindent\normalsize{The strong metal-support interaction (SMSI) leads to substantial changes of the properties of an oxide-supported catalyst after annealing under reducing conditions. The common explanation is the formation of heavily reduced, ultrathin oxide films covering metal particles. This is typically encountered for reducible oxides such as \ce{TiO2} or \ce{Fe3O4}. Zirconia (\zro), a typical catalyst support, is difficult to reduce and therefore no obvious candidate for the SMSI effect. In this work, we use inverse model systems with Rh(111), Pt(111), and Ru(0001) as supports. Contrary to expectations, we show that SMSI is encountered for zirconia. Upon annealing in ultra-high vacuum, oxygen-deficient ultrathin zirconia films ($\approx$\ce{ZrO_{1.5}}) form on all three substrates. However, Zr remains in its preferred charge state of 4+, as electrons are transferred to the underlying metal. At high temperatures, the stability of the ultrathin zirconia films depends on whether alloying of Zr and the substrate metal occurs. The SMSI effect is reversible; the ultrathin suboxide films can be removed by annealing in oxygen.} \\

\end{tabular}

 \end{@twocolumnfalse} \vspace{0.6cm}

  ]

%%%FONT SETUP - please do not change any commands within this section
\renewcommand*\rmdefault{bch}\normalfont\upshape
\rmfamily
\section*{}
\vspace{-1cm}

%%%FOOTNOTES%%%

\footnotetext{\textit{$^{a}$~Institute of Applied Physics, TU Wien, 1040 Vienna, Austria.}}
\footnotetext{\textit{$^{\ddag}$~Present address: Center for Nanomaterials and Chemical Reactions, Institute for Basic Science (IBS) Daejeon 305-701, South Korea.}}
\footnotetext{\textit{$\ast$~Fax: +43 1 58801 13499; Tel: +43 1 58801 13401; E-mail: schmid@iap.tuwien.ac.at}}

%Please use \dag to cite the ESI in the main text of the article.
%If you article does not have ESI please remove the the \dag symbol from the title and the footnotetext below.
\footnotetext{\dag~Electronic Supplementary Information (ESI) available. See DOI: 00.0000/00000000.}

%%%END OF FOOTNOTES%%%

%%%%%%%%%%%%%%%%%%%%%%%%%%%%%%%%%%%%%%%%%%%%%%%%%%%%%%%%%%%%%%%%%%%%%
%% START OF MAIN TEXT
%%%%%%%%%%%%%%%%%%%%%%%%%%%%%%%%%%%%%%%%%%%%%%%%%%%%%%%%%%%%%%%%%%%%%

%\def\imagedir{../../Figures/}
\def\imagedir{.}
\setchemformula{kroeger-vink}

\section{Introduction}
Already in the late 1970es, Tauster \latin{et al.}\ reported a strong change in reactivity after annealing oxide-supported catalysts under reducing conditions.\cite{tauster_strong_1981, tauster_strong_1978, tauster_strong_1978-1} This increase or decrease of reactivity, depending on the reaction, seemed to stem from an interaction between metal particles and their oxide support, hence the effect was named ``strong metal-support interaction'' (SMSI). The effect is reversible; the original state can be recovered by reoxidation. The SMSI effect was studied intensively, as the change in reactivities can be used for tuning the selectivity of oxide-supported catalysts towards the desired end product.\cite{diebold_surface_2003} For the prototypical oxide support \ce{TiO2}, it was shown later that the SMSI effect was due to a heavily reduced oxide film (\ce{TiO_{1.1}}) encapsulating Pt clusters. \cite{datye_comparison_1995, dulub_imaging_2000} This explanation, probably first considered by Meriaudeau \latin{et al.},\cite{meriaudeau_further_1982} was also applied to many other combinations of reducible oxides and metals, e.g.\ Pd/\ce{TiO2}, \cite{bennett_surface_2002} Fe/\ce{TiO2},\cite{pan_encapsulation_1993} Pt/\ce{Fe3O4},\cite{qin_encapsulation_2008} and Pt, Pd, Rh/\ce{CeO2}.\cite{bernal_recent_1999}
This mechanism is very different from the original idea of a modification of the metal's electronic structure by the oxide support, which had led to the term ``SMSI''.\cite{haller_metalsupport_1989} Nevertheless, the term SMSI is still used for this phenomenon.
Early SMSI studies of Ir on different oxide supports found that the tendency of the system to exhibit the SMSI effect depends on the reducibility of the support.\cite{tauster_strong_1978} For materials commonly seen as hard to reduce or non-reducible, such as \ce{HfO2} and \zro{}, no effect was found that went beyond cluster agglomeration. This is in agreement with the explanation of the SMSI effect as covering the metal by a reduced oxide film (suboxide). Only later it was shown that SMSI can also be encountered for Rh/\zro{},\cite{dallagnol_hydrogenation_1985} Pt/\zro{},\cite{hoang_effect_1994, hoang_hydrogen_1995, bitter_state_1997} and Au/\zro{}.\cite{widmann_support_2010} In view of the fact that ZrO$_{2-x}$ suboxides are unstable or at best marginally stable,\cite{abriata_ozrphasediagram_1986, xue_prediction_2013} the question arises whether the accepted mechanism of the metal being coated by a suboxide is also responsible for the SMSI effect on zirconia.

In this work, based on inverse model systems of zirconia on Rh(111) (used for most of this work), Pt(111), and Ru(0001), we show that reducing conditions indeed lead to the formation of oxygen-deficient ultrathin zirconia films covering the metal, although Zr remains in its oxidized, 4+ charge state. The ultrathin films can be removed by oxidation. The influence of the substrate on the growth behaviour is studied in detail. 

\section{Experimental Methods}

The ultra-high vacuum (UHV) system used in this work is a two-vessel setup consisting of seperate chambers for preparation and analysis. The preparation chamber ($p_\mathrm{base}$\,<\,\p{1}{10}) features a sputter gun and heating possibilities to clean the substrate single crystals. Furthermore, a home-built, UHV-compatible sputter source \cite{lackner_SputterSource_2017} for the deposition of Zr is mounted in the chamber. 
The analysis chamber ($p_\mathrm{base}$\,<\,\p{7}{11}) features an Omicron microSTM for scanning tunneling microscopy (STM) at room temperature as well as a SPECS Phoibos 100 analyzer, which was used for x-ray photoelectron spectroscopy (XPS, emission 15\grad{} off-normal) in combination with a lab x-ray source (non-monochromatized Mg K$\alpha$).
The whole system is suspended on springs for vibration damping. For STM we used etched W tips, which were cleaned by pulsing on a Au(110) crystal, and by sputtering. Atomically resolved STM images in this work are corrected for piezo drift as described in Ref.\ \citenum{choi_growth_2014} which gives accurate distance measurements.

We used inverted model systems of supported catalysts on \zro{}. The reason for this is that zirconia has a band gap of 5--6\,eV \cite{french_experimental_1994} and is therefore a prefect insulator, precluding the use of STM on bulk material and complicating XPS studies due to charging effects. Few-monolayer-thick films of zirconia can be studied by STM, as shown by Maurice et al.\ \cite{maurice_epitaxial_1990} and Meinel \latin{et al.}\cite{meinel_surface_2006} Rh(111) was chosen as the main substrate as its lattice parameter fits to zirconia with a ratio of 4:3, resulting in zirconia films with well-defined crystallography. \cite{lackner_surfacestructure_2019} The substrate single crystals, Rh(111), Pt(111), and Ru(0001), have a diameter of $\approx$\,7--9\,mm and a thickness of 2\,mm. After a standard cleaning procedure of several cycles of sputtering (2\,keV, $I_\mathrm{t}$\,=\,3.6\,$\mu$A/cm$^2$) and annealing ($T$\,=\,850 to 950\gradC), Zr was deposited in Ar and \ox{} background (\pox{}\,$\approx$\,\p{5}{7}, $p_\mathrm{Ar}$\,$\approx$\,\p{1}{5} in the UHV chamber) using the sputter source. \cite{lackner_SputterSource_2017} The application of this source is beneficial as Zr is difficult to evaporate due to its low vapor pressure even near the melting point (2128\,K), and the sputter-deposited films exhibit better stability than films deposited by evaporation. \cite{lackner_SputterSource_2017} We define one \zro{} monolayer (ML) as one repeat unit of cubic \zro{}(111), or tetragonal \zro{}(101), with a thickness of $\approx$\,0.3\,nm. 
Sample temperatures were controlled via a thermocouple attached to the sample holder, rather than to the sample plate or the sample directly. This leads to a systematic error, which was corrected by additional measurements with a disappearing-filament pyrometer. We estimate that the temperature values in this work are accurate within $\pm$\,30\gradC{}. Samples were annealed for 10\,min unless noted otherwise. 

\subsection{Zirconia on Rh(111)}
\begin{figure*}[htb!]
\begin{center}
\includegraphics[width=(0.9\hsize)]{\imagedir/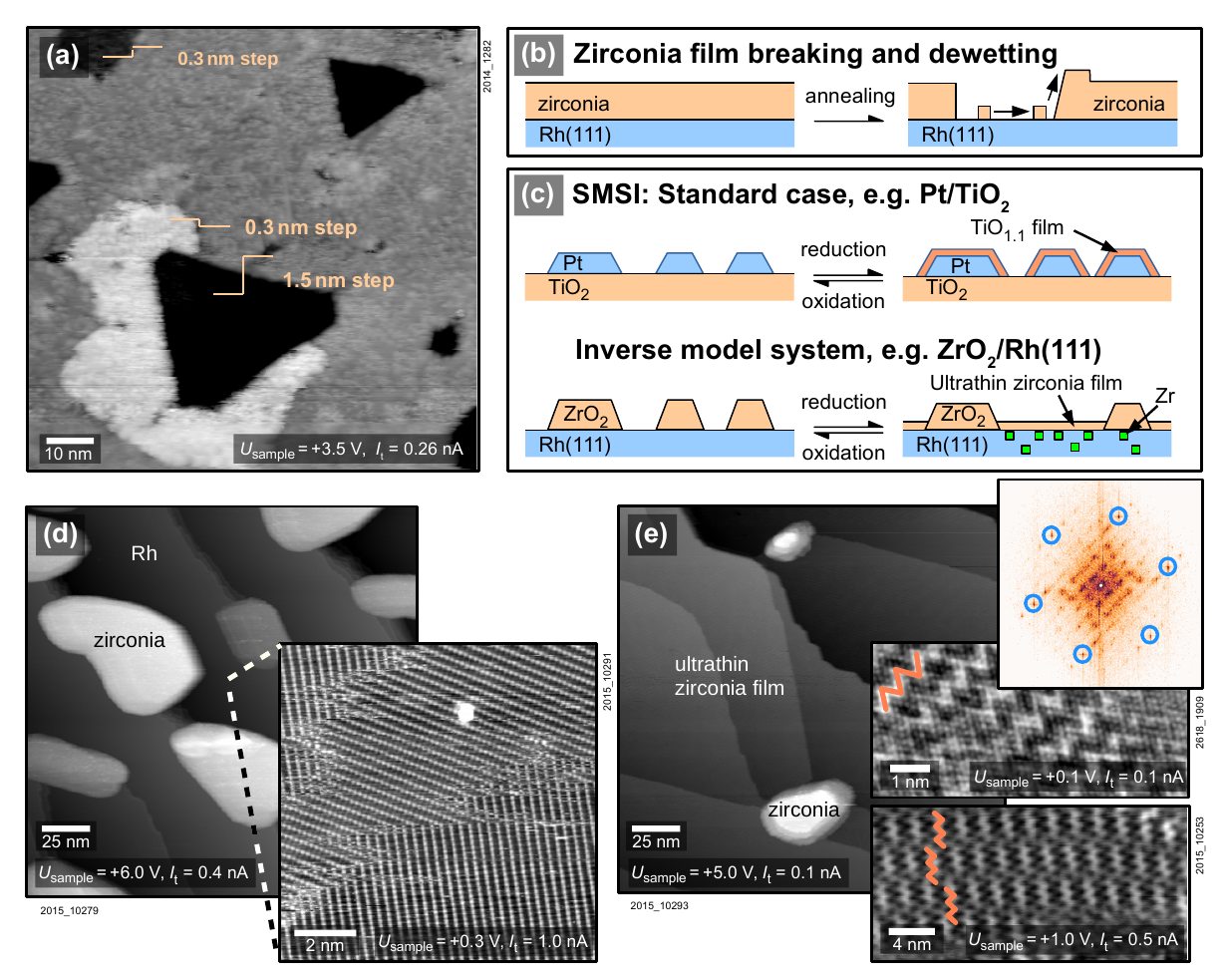}
\caption{The inverse model catalyst \zro /Rh(111). \f(a) \n STM image of a 5\,ML (1.5\,nm)-thick \zro{} film on Rh(111) after annealing at 750\gradC{} in \ox{} shows the beginning of dewetting, as sketched in  \f(b)\n. A few holes reach down to the Rh substrate, and the expelled oxide migrates onto the film.
\f(c) \n The standard SMSI mechanism leading to metal particles being overgrown by substoichiometric oxide films, as well as the mechanism on inverse model systems.
\f(d) \n 2\,ML of zirconia/Rh(111) after annealing at 870\gradC{} in \ox{}: Zirconia dewets the substrate and forms islands. (inset) Rh(111)-\twobyone{}-O superstructure from cooling in oxygen. \f(e) \n After annealing at 870\gradC{} in UHV, less zirconia is contained in islands, and the surface is covered with an ultrathin zirconia film. The STM insets show the ultrathin film with a zigzag moir{\'e}-pattern (marked in orange) in a usual resolution (lower), and high resolution (middle). The Fourier transform (top) clearly shows a 0.35\,nm periodicity (blue circles).}
\label{fig:SMSI_RhSTM}
\end{center}
\end{figure*}

The preparation of an inverse model system of zirconia on a metal starts with the sputter-deposition of a closed zirconia film on a clean Rh(111) single crystal. Upon annealing a 5\,ML \zro{} film in \ox{} ($T >$ 750\gradC{}, usually \pox{}\,=\,\p{5}{7}), zirconia begins to dewet the surface, see Figure \ref{fig:SMSI_RhSTM}a, b. \zro{} migrates to the top of the film, locally increasing the total height of the film by one layer ($\approx$\,0.3\,nm), and the Rh surface gets exposed.

To increase the free Rh surface, and thereby increase the area where the SMSI effect can be studied, only two monolayers (ML) of zirconia were deposited on the substrate. 
The sample was oxidized at a pressure of \p{5}{7} at 870\gradC{}. During this annealing step, zirconia forms islands and reveals the substrate in-between the islands, see Figure \ref{fig:SMSI_RhSTM}d. The substrate either shows a Rh(111) \onebyone{} structure or a \twobyone{}-O superstructure, depending on the oxygen pressure during cooling, see below. The \twobyone{}-O superstructure is seen in the inset of Figure \ref{fig:SMSI_RhSTM}d. (A study of the surface of a mildly annealed 2\,ML-thick film is found elsewhere. \cite{lackner_surfacestructure_2019}) 
When exposing the sample to reducing conditions by annealing at 870\gradC{} in UHV instead of oxygen, an ultrathin film is formed that covers the Rh surface completely, see Figure \ref{fig:SMSI_RhSTM}e. The total amount of zirconia in the islands decreases drastically. The remaining islands cover only $\approx 2$\% of the surface with an average height of about 5\,ML, thus they accommodate only $\approx$\,5\% of the material deposited. The ultrathin film between the islands can be assumed to be one layer of zirconia(111), the remaining Zr must be dissolved in the Rh substrate (see below). The process is reversible; the ultrathin zirconia film disappears upon annealing in oxygen, the \zro{} islands grow in size, using both Zr in the ultrathin film and Zr dissolved in the metal (see below). The total amount of zirconia on the sample decreases with each reduction-oxidation cycle, as some Zr is lost to the bulk, see below. For an initial deposition of 2\,ML, 10\% of the total Zr is lost after the first reduction-oxidation cycle. The reduction-oxidation cycle is sketched in Figure \ref{fig:SMSI_RhSTM}c, both for inverse model systems and for real catalysts.

A closer look at the ultrathin film reveals a hexagonal lattice with interatomic distances of 0.35\,nm, as is typical for ultrathin zirconia films, \cite{choi_growth_2014, antlanger_pt3zr0001_2012} see the Fourier transform (FFT) in Figure \ref{fig:SMSI_RhSTM}e. When comparing two domains rotated by a multiple of $\approx$\,60\grad{}, their lattices agree within 1\%, demonstrating that the deviations from an exactly hexagonal structure are small. The lattice constant of 0.35\,nm is also confirmed by LEED (not shown), when using a tetragonal zirconia film and the Rh(111) lattice as a references. STM images without atomic resolution mainly show a moir{\'e} pattern, typically with a zigzag appearance (insets of Figure \ref{fig:SMSI_RhSTM}e), which is however gradually lost when annealing at higher temperatures (for details see Figure S1 in the Supporting Information (SI)).

After annealing in \pox{}\,=\,\p{5}{7}, a \twobyone{}-O superstructure as in the inset of Figure \ref{fig:SMSI_RhSTM}d can form on the Rh(111) surface between the multilayer zirconia islands. Whether or not the superstructure forms depends on the oxygen pressure \pox{} (or chemical potential \muox{}) during cool-down. \cite{ganduglia-pirovano_structural_1999, kohler_high-coverage_2004} To test whether the disappearance of the ultrathin zirconia film upon annealing in oxygen is influenced by oxygen adsorption on the Rh(111) surface, the experiment was repeated with a different \muox{} during cooling. Instead of keeping constant \pox{}$ = $\p{5}{7} during cooling down to $\approx$\,300\gradC{}, the sample was cooled from 870\gradC{} to $\approx$\,730\gradC{} while keeping the chemical potential of oxygen constant at \muox{}\,=\,$-2.3$\,eV, where the coverage of oxygen on Rh(111) should be very low. \cite{ganduglia-pirovano_structural_1999, kohler_high-coverage_2004} Below 730\gradC{}, where a pressure of $p <$\,\p{1}{9} was reached, no more oxygen was supplied to the chamber. At this pressure, the impingement rate is low enough to have no effect on the film formation. The resulting surface was similar to the one cooled in \ox{}, as zirconia islands still formed and the ultrathin film was removed. Between the islands,  however, the bare Rh(111) substrate was observed instead of the \twobyone{}-O superstructure (not shown). In all other aspects, the result was indistinguishable from one found while cooling in \ox{}, e.g.\ subsequent annealing in UHV led to the formation of an ultrathin zirconia film.

\subsection{Zirconia on Pt(111) and Ru(0001)}
\label{Ssec:PtRuSTM}

\begin{figure*}[htb!]
\begin{center}
\includegraphics[width=(0.8\hsize)]{\imagedir/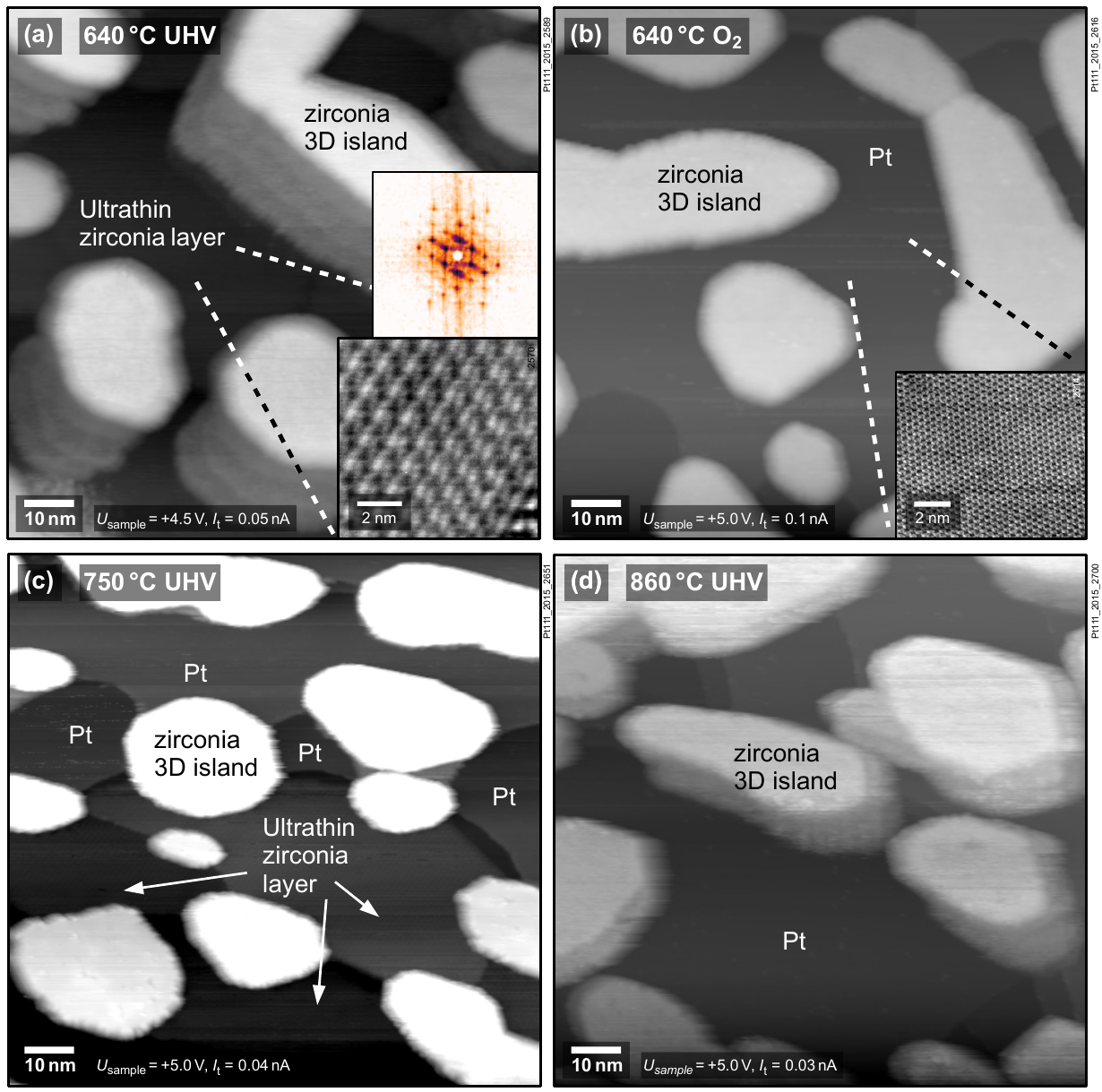}
\caption{The SMSI effect of zirconia on Pt(111): \f(a) \n Annealing at 640\gradC{} in UHV after deposition yields an ultrathin film between islands. The inset shows an STM image of the \sn{} moir{\'e} pattern of the ultrathin film and its FFT. \f (b) \n The ultrathin film is completely removed by annealing at 640\gradC{} in \ox{}; the inset shows the Pt(111) lattice. \f (c) \n An ultrathin zirconia film, partially covering the Pt, forms again by annealing at 750\gradC{} in UHV, but is removed \f (d) \n when annealing at 860\gradC{} in UHV. The islands in (a) and (c) show artifacts (``shadows'') due to an imperfect STM tip.}
\label{fig:SMSI_PtSTM}
\end{center}
\end{figure*}

To examine whether the observed phenomena are specific to the Rh(111) substrate, we have studied \zro{} films on two other metals, Pt(111) and Ru(0001). Figure \ref{fig:SMSI_PtSTM}a shows a zirconia film of $\approx$\,5\,ML thickness deposited on Pt(111) and annealed at 640\gradC{} in UHV directly after sputter deposition, leading to the formation of zirconia islands and an ultrathin zirconia film in between. The ultrathin zirconia film shows a Zr--Zr distance of 0.350 $\pm$  0.003\,nm, as also observed for \ptzr{}(0001) \cite{antlanger_pt3zr0001_2012} and \ce{Pd3Zr}(0001). \cite{choi_growth_2014} The moir{\'e} pattern shown in the inset of Figure \ref{fig:SMSI_PtSTM}a exhibits the same \sn{} superstructure (w.r.t.\ the substrate) as ultrathin zirconia films on Pt-terminated \ptzr{}(0001). \cite{antlanger_pt3zr0001_2012} The creation of the ultrathin film can be reversed by annealing at 640\gradC{} in \p{5}{7} \ox{}, see Figure \ref{fig:SMSI_PtSTM}b.

The following experiments show that the formation of ultrathin films on Pt(111) depends on the preparation conditions and film thickness, in contrast to Rh(111). 
When reannealing the oxidized film at 640\gradC{} in UHV, no ultrathin film forms. Only upon annealing at 750\gradC{} in UHV, $\approx 1/4$ of the Pt(111) surface is covered with an ultrathin zirconia film, while $\approx 1/3$ of the surface is still bare Pt, see Figure \ref{fig:SMSI_PtSTM}c. The same film was annealed in \pox{}\,=\,\p{5}{7} at 640\gradC{} (which removes the ultrathin film), followed by annealing in UHV at 860\gradC{}; the surface remains covered by \zro{} islands, yet the ultrathin film is not found, although increasing the temperature at a constant (though negligible) \ox{} pressure corresponds to more reducing conditions. 
In a second experiment, an ultrathin film could also be produced by annealing 2\,ML of \zro{} on Pt(111) at 640\gradC{} in \p{5}{7} \ox{} followed by 30\,min of UHV annealing at the same temperature (not shown). 

\begin{figure*}[htb!]
\begin{center}
\includegraphics[width=(0.8\hsize)]{\imagedir/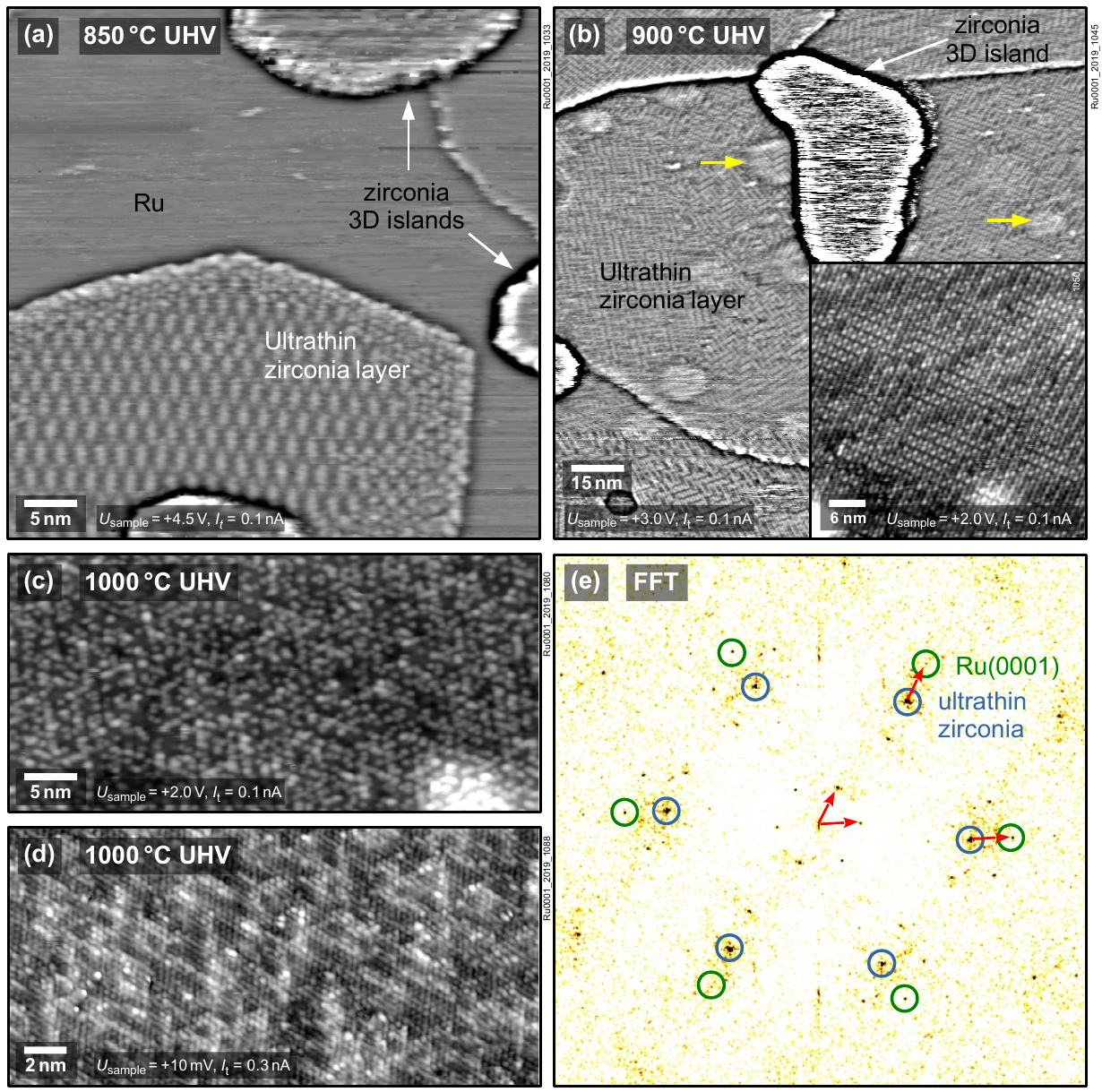}
\caption{Growth of ultrathin zirconia films on Ru(0001): \f(a) \n After annealing 1.5\,ML of \zro{} at 850\gradC{} in UHV, small patches of ultrathin zirconia form around islands. As similar zigzag pattern as on Rh(111) is formed. \f (b) \n After UHV-annealing at 900\gradC{}, the ultrathin film fully wets the substrate. (The bright round areas (yellow arrows) originate from implanted Ar bubbles typical for Ru. \cite{schmid_quantum_1996, gsell_effect_1998}) \f (inset) \n A small-area image reveals a complex structure with a base hexagonal pattern overlaid by rows of bright species. \f (c) \n After UHV-annealing at 1000\gradC{}, high-bias images show a seemingly disordered pattern, while \f (d) \n low-bias images reveal the film to still exhibit the $\approx 0.35$\,nm periodicity expected from ultrathin zirconia films, as can be seen in the FFT \f (e)\n. By resolving the lattices of both Ru(0001) and the ultrathin zirconia film, the moir{\'e} vectors can be found. Images (a) and (b) are high-pass filtered.}
\label{fig:SMSI_RuSTM}
\end{center}
\end{figure*}

SMSI zirconia films on Ru(0001) grow similarly to Rh(111), in contrast to Pt(111): After annealing 1.5\,ML of \zro{} on Ru(0001) at 950\gradC{} in \pox{}\,=\,\p{5}{7}, islands form. UHV-annealing at 850\gradC{} leads to patches of ultrathin zirconia around islands, see Figure \ref{fig:SMSI_RuSTM}a. These patches show a similar zigzag pattern as on Rh(111). Finally, after annealing at $T$\,=\,900\gradC{} in UHV, the ultrathin zirconia fully covers the metal substrate (and, as for Rh(111), the zigzag pattern disappears). This ultrathin film has an underlying hexagonal pattern with rows of bright features on top, see inset of Figure \ref{fig:SMSI_RuSTM}b. Even at 1000\gradC{}, the ultrathin film is not removed, in stark contrast to Pt(111). Instead, the film shows disordered features when measured with high STM bias ($V_\mathrm{sample} = +2$\,V), see Figure \ref{fig:SMSI_RuSTM}c. However, at low STM bias ($V_\mathrm{sample} = +0.01$\,V), an ordered structure is again resolved, see Figure \ref{fig:SMSI_RuSTM}d. In the FFT (see Figure \ref{fig:SMSI_RuSTM}e), both the Ru(0001) lattice and the typical lattice of ultrathin zirconia films are resolved, alongside the resulting moir{\'e} pattern. It comes as a surprise that even at such high temperatures, and in presence of the disordered features, the underlying periodicity of the film remains intact, although the lattice constant of the ultrathin film is somewhat smaller than usual (0.344 instead of 0.35\,nm).

\subsection{Photoelectron Spectroscopy}
\label{Ssec:SMSI_XPS}
\begin{figure*}[htb!]
\begin{center}
\includegraphics[width=(0.8\hsize)]{\imagedir/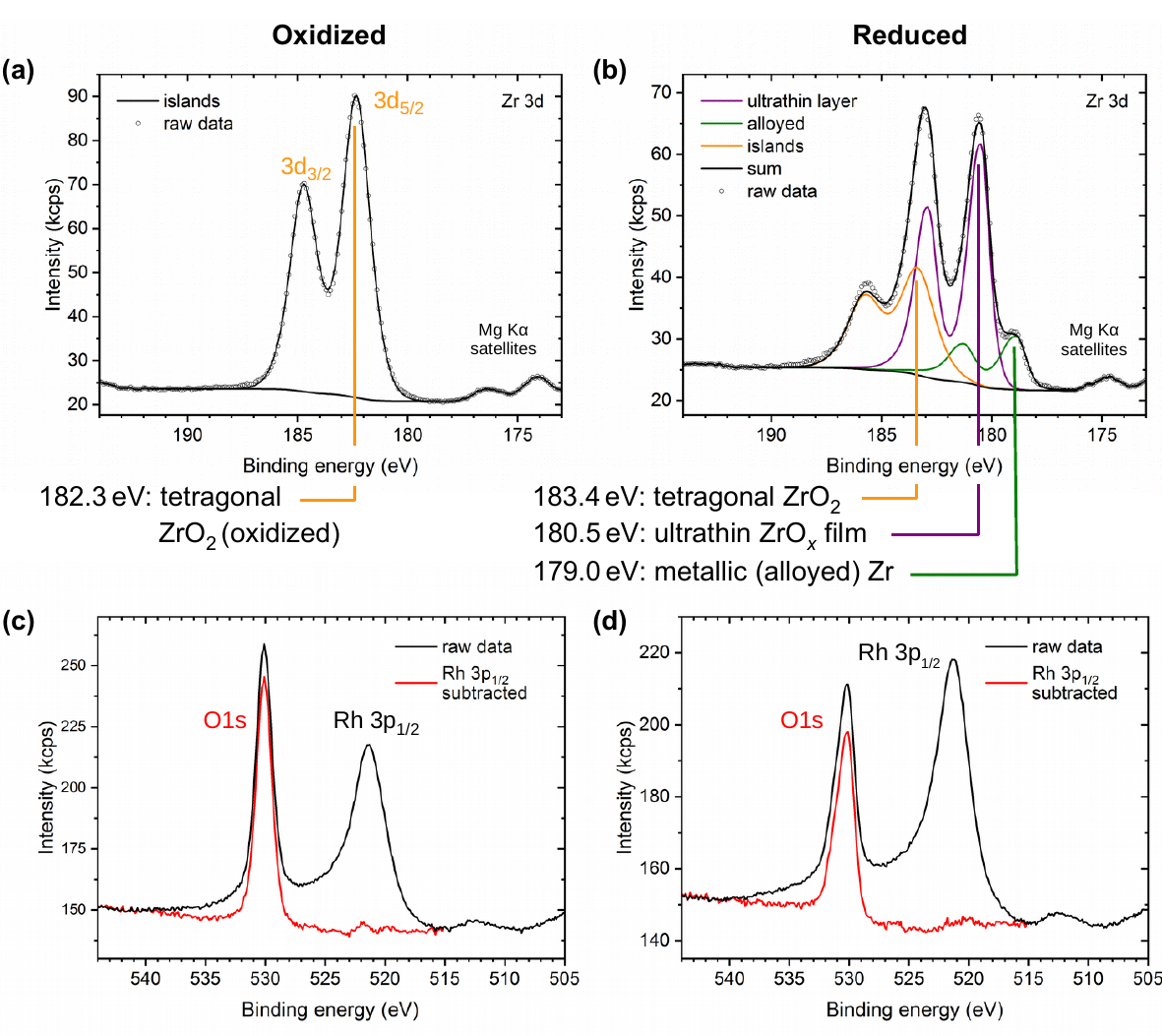}
\caption{\n XPS of oxidized \f (a,c) \n and reduced \f(b,d) \n 2\,ML zirconia films on Rh. \f(top) \n Zr 3d region. The oxidized preparation shows one doublet from bulk-like \zro{} islands. The reduced preparation shows a shifted island doublet and two new doublets assigned to the ultrathin film and to metallic Zr, respectively. \f(bottom) \n O 1s and Rh 3p region, showing both, the raw data and the data after subtraction of the Rh 3p peak.}
\label{fig:SMSI_XPS}
\end{center}
\end{figure*}

Films of 2\,ML of zirconia/Rh(111) were investigated using XPS in both the reduced and the oxidized state, see Figure \ref{fig:SMSI_XPS}. The oxidized system ($T$\,=\,820\gradC{}, \pox{}\,=\,\p{5}{7}) shows only one doublet in the Zr 3d region with a binding energy \EB{} of Zr 3d$_{5/2}$ = 182.3\,eV, which is identified as essentially fully oxidized, tetragonal zirconia. \cite{lackner_electronic_2019} This comes to no surprise as STM shows that all zirconia is contained in islands, see Figure \ref{fig:SMSI_RhSTM}d, while the Rh(111) surface is exposed and acts as an oxygen dissociation catalyst; \cite{lackner_electronic_2019} thus, annealing in oxygen leads to full oxidation of zirconia. On the other hand, three doublets are found for the reduced system ($T$\,=\,820\gradC{} in UHV). The first can again be attributed to zirconia islands (183.4\,eV), although the signal is shifted by 1.1\,eV towards higher \EB{}, caused by reduction (n-type doping by oxygen vacancies; the Fermi level is closer to the conduction band). \cite{lackner_electronic_2019} The full width at half maximum (FWHM) of the peaks in this doublet increases from 1.47\,eV to 1.84\,eV; possibly due variations of the doping level (slight non-stoichiometry) between different islands. The second doublet originates from ultrathin zirconia (180.5\,eV), as previously observed on \ptzr(0001). \cite{li_growth_2015} The third doublet (179.0\,eV) is slightly shifted with respect to metallic Zr (178.6\,eV \cite{bakradze_Zr_2011}). Such a shift is typical when alloying occurs, in this case with the Rh substrate. (In the \ptzr{} intermetallic phase, the Zr 3d$_{5/2}$ peak is shifted more, to 179.6\,eV. \cite{li_forward-focusing_1993}) The fact that \zro{} can be reduced to its metallic state on a metallic substrate was already shown for \zro{}/Pd. \cite{guo_x-ray_1999, kopfle_zirconium-palladium_2017}

The area ratio of the various Zr 3d doublets strongly depends on the preparation conditions. 
The peak area of the tetragonal \zro{} islands depends on the amount of \zro{} deposited. The alloy peak area depends on the annealing temperature, annealing time, and on the amount of zirconia available for reduction. It can be both higher and lower than in the spectra shown in Figure \ref{fig:SMSI_XPS}b; in the case of very little deposited \zro{} (e.g.\ 1.1\,ML or 1.2\,ML, see below), the peak vanishes below the detection limit, which is $\approx$\,0.04\,ML.

The O 1s region shows a single peak for both reducing and oxidizing preparation conditions, overlapping with the tail of the Rh 3p$_{1/2}$ substrate peak (\EB{} = 521.3\,eV), see Figure \ref{fig:SMSI_XPS}c, d. By subtracting a normalized Rh 3p$_{1/2}$ peak measured on a clean Rh(111) surface, the O 1s peak can be isolated. The O 1s peak of the oxidized preparation is found at 530.1\,eV with a FWHM value of 1.58\,eV, as for oxidized tetragonal zirconia. \cite{lackner_electronic_2019}
In the reduced preparation, a high-binding-energy shoulder appears, increasing the total FWHM to 1.83\,eV. The peak maximum stays nearly constant at 530.2\,eV. This is expected for a system consisting of an ultrathin zirconia film with a lower \EB{} (529.9\,eV \cite{li_growth_2015}) and islands with a higher \EB{} (due to slight reduction, i.e., n-doping, the peak of tetragonal \zro{} shifts to \EB{} $\approx$\,531.0\,eV \cite{lackner_electronic_2019}). The resolution of a lab-based XPS setup is insufficient for accurate deconvolution of these O 1s signals. Thus, the combined intensities lead to a peak broadening, yet only a small shift of the peak maximum. 

On Pt(111), no peak for metallic (alloyed) Zr was found. This may be partly due to the fact that Zr alloyed with Pt is shifted towards substantially higher \EB{} (179.6\,eV\cite{li_growth_2015}, +0.6\,eV w.r.t.\ Zr alloyed with Rh), and therefore overlaps more with the ultrathin zirconia peak. Thus, only higher amounts ($>$\,0.08\,ML) would be detectable.

On Ru(0001), 1.5\,ML of \zro{} were first annealed at 950\gradC{} in \ox{} and then annealed step-wise in UHV, as already described in section \ref{Ssec:PtRuSTM}. A metallic (alloyed) Zr peak appeared only at annealing temperatures $\geq$\,950\gradC{}. Even after annealing at 1020\gradC{} for 25\,min in UHV, the metallic peak remained small ($\lesssim$\,15\% of the total Zr 3d peak area). Additionally, the peak assigned to the ultrathin film shifted with higher $T$; at 900\gradC{}, where the film first covers the whole substrate, Zr 3d$_{5/2}$ lies at 180.6\,eV. It shifts by $-0.2$\,eV (towards lower \EB{}) after annealing at 950\gradC{} and by another $-0.1$\,eV after annealing at 1020\gradC{} for 25\,min.

\subsection{Stoichiometry of Ultrathin Zirconia}
\label{Ssec:SMSI_Stoichiometry}
The area ratio of Zr 3d (excluding alloyed Zr) to O 1s can be used to calculate the stoichiometry of ultrathin films. As a standard for XPS quantification, we used a closed, fully oxidized (using Rh clusters as catalyst, see Ref.\ \citenum{lackner_electronic_2019}) 5\,ML-thick \zro{} film annealed at 610\gradC{} (below the dewetting temperature). The surface structure of this film is known. \cite{lackner_surfacestructure_2019}
When comparing films of different thickness such as the 5\,ML standard and the ultrathin film, we have to account for the different attenuation of the Zr 3d and O 1s signals. We have therefore simulated photoelectron transport in both systems using the SESSA code, \cite{smekal_simulation_2005} which then allows us to determine the stoichiometry of the ultrathin zirconia film. This approach leads to a number of uncertainties, which are discussed in detail in Chapter 2 in the SI. This method yields a stoichiometry of $\mathrm{ZrO}_{{1.0}_{-0.12}^{+0.26}}$ for the ultrathin SMSI film on Rh(111). For the zirconia islands of the oxidized preparation, the result is \ce{ZrO_{1.94$\pm$0.14}}; the expected value of \zro{} lies within the range.
The same analysis was conducted for ultrathin zirconia films on Ru(0001) at different temperatures, revealing a similar result as on Rh(111): The analysis gives a stoichiometry of \ce{ZrO_{1.07}} after annealing at 900\gradC{} in UHV, \ce{ZrO_{1.01}} at 950\gradC, \ce{ZrO_{1.11}} at 1000\gradC{} and finally \ce{ZrO_{1.04}} after annealing at 1020\gradC{} for 25\,min. Differences between these values are within the error bars mentioned above.

Another method to gain information on the stoichiometry is the direct comparison of the O 1s intensity of an ultrathin zirconia film with an O--Rh--O trilayer, \cite{gustafson_self-limited_2004} both prepared on the same Rh(111) single crystal. The O--Rh--O trilayer was prepared by annealing Rh(111) at $T=410$\gradC{} in \pox{}\,=\,\p{1.5}{4} (using an oxygen doser similarly shaped as a shower head; the chamber pressure was \p{5}{6}). In this pressure regime, the formation of a surface oxide is self-limiting and no 3D oxide islands are formed. \cite{gustafson_self-limited_2004} To minimize the amount of remaining 3D \zro{} islands after the preparation of the ultrathin SMSI films, two ultrathin zirconia films were prepared with only 1.2\,ML and 1.1\,ML of zirconia, respectively. These zirconia films were annealed in oxygen at $T=550$\gradC{} and 670\gradC{}, respectively, to gain fully oxidized islands, then reduced for 20\,min at $T=$950\gradC{} and 70\,min at $T=860$\gradC{}, respectively, in UHV. To compensate for possible variations of the x-ray intensity, the x-ray-induced sample current was measured at the sample holder before inserting the sample; the results were normalized by this value. By this direct comparison method, inaccuracies induced by simulations and reference films can be avoided. However, it has to be assumed that no oxygen was dissolved in the Rh substrate; especially for the \ce{RhO2} film, this might not be true, and would lead to underestimating of the zirconia oxygen content. Furthermore, the area of uncovered substrate must be estimated from (local) STM images. The resulting O 1s intensity ratios between the zirconia-covered surface and the \ce{RhO2} film are 0.62 for the 1.2\,ML and 0.50 for the 1.1\,ML zirconia deposition. A ratio of 0.75 is expected for a fully oxidized trilayer of \zro{} due to the larger lattice constant (0.35\,nm for zirconia as compared to 0.302\,nm for O--Rh--O \cite{gustafson_self-limited_2004}). The resulting stoichiometries are therefore \ce{ZrO_{1.7}} and \ce{ZrO_{1.4}}, respectively. Comparing the photoelectron-induced O$_\mathrm{KLL}$ Auger peaks yields \ce{ZrO_{1.6}} and \ce{ZrO_{1.4}}, respectively. Using the Auger peaks is, on the one hand, less accurate than using O 1s due to the lower intensity of Auger peaks. On the other hand, Auger peaks have a higher surface sensitivity, i.e.\ are less sensitive to O dissolved in the Rh bulk.

\begin{table*}
\renewcommand{\arraystretch}{1.4}
%\makebox[\textwidth][c]{
\begin{center}
\begin{tabular}{c c c c c c c}
\hline 
Substrate & Growth & Method & Standard & Source &O:Zr ultra-thin film & Assumptions \\ 
\hline 
Rh(111) & SMSI & XPS & \ce{RhO2} & this work & $\approx$\,\ce{ZrO_{1.5}} & Islands: \zro{} \\ 
%\hline 
Rh(111) & SMSI & XPS & 5\,ML \zro{} & this work & $\mathrm{ZrO}_{{1.0}_{-0.12}^{+0.26}}$ & Islands: \zro{}$^a$ \\
Ru(0001) & SMSI & XPS & 5\,ML \zro{} & this work & $\approx \mathrm{ZrO}_{1.1}$ & Islands: \zro{} \\
%\hline 
\ptzr{}(0001) & Alloy & XPS & 1\,ML water & Ref.\ \citenum{lackner_PhD_2019} & \ce{ZrO_{1.4}} & Islands: \zro{} \\ 
%\hline 
\ptzr{}(0001) & Alloy & AES & \ce{RhO2} & Ref.\ \citenum{antlanger_pt3zr0001_2012} & \ce{ZrO_{1.62}} & No islands \\ 
%\hline 
\ptzr{}(0001) & Alloy & synchrotron-based XPS & -- & Ref.\ \citenum{li_growth_2015} & \ce{ZrO_{1.82}} & Islands: \ce{ZrO_{1.82}} \\ 
%\hline 
\ce{Pd3Zr}(0001) & Alloy & AES & \ce{RhO2} & Ref.\ \citenum{choi_growth_2014} & \ce{ZrO_{2.19}} & No islands \\ 
\hline 
\end{tabular} 
%}
\end{center}
\begin{flushright}
$^a$ The possibility of reduced islands is included in the error bars.
\end{flushright}
\caption{Stoichiometries of ultrathin zirconia films with different preparation methods and substrates}
\label{Tab:Stoichiometries}
\end{table*}

Taken together, the quantitative XPS measurements indicate a substoichiometric ultrathin film. This implies that other ultrathin zirconia films may also be substoichiometric, regardless of whether they were obtained by oxidation of alloys, \cite{choi_growth_2014, antlanger_pt3zr0001_2012, choi_metal_2016} or deposition of Zr and oxidation. \cite{meinel_surface_2006} In fact, some previous results have indicated substoichiometric films, but this interpretation was attributed to the limited accuracy of the measurement rather than nonstoichiometry. All measured stoichiometries of ultrathin zirconia films are summarized in Table \ref{Tab:Stoichiometries}. Comparison of the Auger signals between the ultrathin zirconia films and a \ce{RhO2} trilayer led to compositions of \ce{ZrO_{1.62}} and \ce{ZrO_{2.19}} for the ultrathin oxides on \ptzr{} \cite{antlanger_pt3zr0001_2012} and \ce{Pd3Zr}, \cite{choi_growth_2014} respectively; the latter value is rather inaccurate due to O dissolved in the \ce{Pd3Zr} bulk. Using a ML of water \cite{lackner2018water} as a reference, we found that ultrathin zirconia on \ptzr{} has a stoichiometry of \ce{ZrO_{1.4}}. \cite{lackner_PhD_2019}
A synchrotron-based XPS study \cite{li_growth_2015} has found \ce{ZrO_{1.82}} for both, the ultrathin oxide and 3D oxide islands on \ptzr{}. As it is unlikely that few-monolayer-thick 3D islands are strongly non-stiochiometric, \cite{lackner_electronic_2019} this result may be also related to inaccurate peak deconvolution and point towards an ultrathin film that contains even less O. It should be noted that not all ultrathin zirconia films necessarily have the same stoichiometry. 

\section{Discussion}
Metal--\zro{} systems clearly show the so-called SMSI effect as observed for reducible oxides such as the prototypical system Pt/\ce{TiO2}. \cite{tauster_strong_1981, datye_comparison_1995, dulub_imaging_2000}
Upon reduction, the metal is covered by an ultrathin oxide film, which is substoichiometric ($\approx$\,\ce{ZrO_{1.5}}), though probably to a lesser degree than for e.g.\ \ce{TiO2}, where the film exhibits a \ce{TiO_{1.1}} stoichiometry. \cite{dulub_imaging_2000}
The ultrathin zirconia films on Rh(111) and Pt(111) have essentially the same lattice constant (0.35\,nm) as the respective films on \ptzr{}(0001) \cite{antlanger_pt3zr0001_2012} and \ce{Pd3Zr}(0001), \cite{choi_growth_2014} so it is likely that they have the same structure, an O--Zr--O trilayer with additional oxygen vacancies. Density functional theory (DFT) calculations show that oxygen vacancies can form much more easily in such a metal-supported ultrathin zirconia film than in bulk zirconia; for oxygen at the oxide-metal interface, the vacancy formation energy is about half the bulk value (2.92\,eV vs.\ $\approx 6$\,eV). \cite{puigdollers_2017}
The reason is that Zr in the ultrathin film can remain in its preferred 4+ state upon formation of an oxygen vacancy; the electrons originally located at the oxygen sites are then transferred to the metal substrate. In addition, if an interface oxygen gets removed, a strong Zr--metal bond can form. \cite{wang_metal-metal_1993}
In contrast to oxides of polyvalent metals, reduction of the ultrathin zirconia films requires transferring two electrons per missing oxygen atom to the metal to circumvent the formation of Zr$^{3+}$, thus the non-stoichiometry will be limited by the electrostatic field induced by the charge transfer.

The calculated vacancy formation energy of 2.92\,eV at the interface of the ultrathin film \cite{puigdollers_2017} roughly agrees with the experimental conditions for forming a complete layer of ultrathin zirconia on Rh; a chemical potential of \muox{}\,$=-2.92$\,eV corresponds to an \ox{} pressure of \p{4}{12} at $T$ = 870\gradC{}. It should be noted, however, that the formation of ultrathin zirconia films on Pt can start already at lower temperatures (observed for 640\gradC{}) --- a fact that points towards zirconia reduction being easier on Pt than on Rh and Ru. This trend is also observed for the reduction of mildly oxidized Zr, see Chapter 3 in the SI. 
One reason for this behaviour 
%However, the different work functions of the substrates [Pt(111): 5.90\,eV (5.92\,eV), Rh(111): 5.16\,eV, Ru(0001): 4.65\,eV (5.37\,eV); calculated patra_properties_2017 (experimental chiarotti_work_1994)] influence the extent of band bending at the zirconia/substrate interface. Assuming that the interface between the film and the substrate is essentially identical for all three substrates, a higher work function and thus stronger band bending leads to easier electron transfer for zirconia reduction.
is the difference in strength of metal--Zr bonds; in case of an oxygen vacancy at the interface, O--Zr bonds can be compensated by metal--Zr bonds. Pt--Zr bonds are stronger than e.g.\ Rh--Zr bonds, as indicated by the alloy formation enthalpies, $-128$\,kJ\,g$^{-1}$\,atom$^{-1}$ for \ce{Pt3Zr}, \cite{srikrishnan_Pt3Zrformeng_1974} vs.\ $-72$\,kJ\,g$^{-1}$\,atom$^{-1}$ for \ce{Rh3Zr}. \cite{LandoltBornstein_phasediagramRhZr}
The strong Pt--Zr bonds facilitate the formation of a reduced zirconia film on Pt.

The substrate also influences the stability of the ultrathin zirconia film at high temperatures. On Pt(111), the zirconia film starts to vanish already after annealing at 750\gradC{} in UHV (or does not cover the whole surface), while on Rh(111) and Ru(0001), the ultrathin film remains stable at far higher $T$. This behaviour can be explained by different diffusion and alloying behaviour of Zr in the substrate materials; diffusion of Zr into the Pt bulk is faster than for Rh at the same temperature, as shown by XPS (Figure S2 and section 3 in the SI). While the ultrathin film can form on any of these substrates (electrons from oxygen vacancies can be transferred to these metals), the competing process under reducing conditions --- complete decomposition of the film and reduction to metallic Zr, which then forms an alloy with the substrate --- starts to dominate at lower temperatures for Pt than for the Rh and Ru substrates.

However, this does not explain the absence of an ultrathin zirconia film after annealing \zro{}/Pt(111) at higher temperatures combined with remaining \zro{} islands. One could envision that all \zro{} islands would be transformed first to reduced ultrathin zirconia (which spreads out over the remaining surface) and would only then be fully reduced to metallic Zr upon annealing at more and more reducing conditions. Before the ultrathin film vanishes, all material contained in islands would be consumed, but this is not the case at least for the Pt substrate.
We therefore conclude that the decomposition of the \zro{} islands is also kinetically hindered. As soon as the \zro{} has decomposed, incorporation of the Zr into the ultrathin (substoichiometric) film and dissolution into the bulk will be competing processes; the branching ratio depends on the temperature and the substrate material. 

For our inverse catalysts, diffusion into the bulk is basically unlimited. This would not be the case for ``real'' catalysts, i.e.\ metal nanoparticles supported by zirconia, where no semi-infinite metal reservoir is present. For the example of Pt nanoparticles on a \zro{} support, Pt would get saturated with Zr; then, formation of an ultrathin zirconia film would occur also at high temperatures, as the competing process of diffusion of all Zr into the Pt would be impossible. On the other hand, Zr dissolution in metal nanoparticles may lead to an increased lattice constant of the metal catalyst, which is not observed in our case (no indications of subsurface misfit dislocations). Since the metal will be covered by the ultrathin zirconia in this state, a modification of the metal lattice constant will not modify the surface chemistry, however. 

Similar to the reducible oxides, the SMSI effect is reversible also for metal--\zro{} systems. We can exclude competition between the ultrathin zirconia and oxygen adsorption on the metal as a driving force for disappearance of the ultrathin zirconia, as demonstrated by cooling at conditions where adsorbed O on Rh should be unstable. Rather, the effect of oxidizing conditions must be seen as the ultrathin suboxide becoming unfavorable with respect to fully oxidized \zro{}.
Under oxidizing conditions, at sufficiently high temperatures, not only the ultrathin substoichiometric film will be converted to \zro, but also dissolved Zr will diffuse, eventually reaching the surface where it reacts with oxygen and is again incorporated in the fully oxidized (bulk-like) \zro{}.

\section{Summary}

We have demonstrated the so-called SMSI effect for inverse model catalysts, zirconia on metal substrates (Rh, Pt, and Ru). When annealed under reducing conditions, the substrate between 3D zirconia islands is covered by a sub-stoichiometric, ultrathin zirconia film similar to the zirconia films previously obtained by oxidation of zirconium alloys. When annealing in oxygen, all Zr becomes fully oxidized and is incorporated in zirconia islands; the ultrathin film disappears. The formation of a substoichimetric oxide is facilitated by contact to a metal, which solves the long standing problem of the SMSI effect observed for oxides that are usually non-reducible, Zr in the substoichiometric films can stay in its preferred 4+ state due to electron transfer to the substrate.

\section*{Conflicts of interest}
There are no conflicts to declare.

\section*{Acknowledgements}
The authors thank Edvin Lundgren from Lund University for kindly lending us a Ru(0001) single crystal. This work was supported by the Austrian Science Fund (FWF) under project number F4505 (Functional Oxide Surfaces and Interfaces -- FOXSI).

%%%%%%%%%%%%%%%%%%%%%%%%%%%%%%%%%%%%%%%%%%%%%%%%%%%%%%%%%%%%%%%%%%%%%
%% Bibliography
%%%%%%%%%%%%%%%%%%%%%%%%%%%%%%%%%%%%%%%%%%%%%%%%%%%%%%%%%%%%%%%%%%%%%
\bibliography{SMSI_bibliography}
\bibliographystyle{rsc} %the RSC's .bst file

\end{document}